\documentclass{PoS}
\usepackage{graphics,psfrag}
\usepackage{graphicx,psfrag}

\title{Progress on charm semileptonic form factors\\ from 2+1 flavor lattice QCD}

\ShortTitle{Progress on $D\to K(\pi)l\nu$ form factors}

\newcommand{\mb}[1]{\mathbf{#1}}
\newcommand{\mc}[1]{\mathcal{#1}}
\newcommand{\mr}[1]{\mathrm{#1}}
\newcommand{\bra}[1]{\langle#1|}
\newcommand{\ket}[1]{|#1\rangle}

\author{%
\speaker{Jon~A.~Bailey}$^a$\thanks{jabailey@fnal.gov},
A.~Bazavov$^b$,
C.~Bernard$^c$,
C.~Bouchard$^e$,
C.~DeTar$^d$,
A.X.~El-Khadra$^e$,
E.D.~Freeland$^c$,
W.~Freeman$^b$,
E.~Gamiz$^{a,e}$,
Steven~Gottlieb$^{e,f,g}$,
U.M.~Heller$^h$,
J.E.~Hetrick$^i$,
A.S.~Kronfeld$^a$,
J.~Laiho$^c$,
L.~Levkova$^d$,
P.B.~Mackenzie$^a$,
M.B.~Oktay$^d$,
M.~Di Pierro$^j$,
J.N.~Simone$^a$,
R.~Sugar$^k$,
D.~Toussaint$^b$,
and
R.S.~Van~de~Water$^l$ \\ \\
\llap{$^a$}Theoretical Physics Department, Fermilab, Batavia, IL  60510, USA \\
\llap{$^b$}Department of Physics, University of Arizona, Tucson, AZ  85721, USA \\
\llap{$^c$}Department of Physics, Washington University, St.~Louis, MO  63130, USA \\
\llap{$^d$}Physics Department, University of Utah, Salt Lake City, UT  84112, USA \\
\llap{$^e$}Physics Department, University of Illinois, Urbana, IL  61801, USA \\
\llap{$^f$}Department of Physics, Indiana University, Bloomington, IN  47405, USA \\
\llap{$^g$}National Center for Supercomputing Applications, University of Illinois, Urbana, IL  61801, USA \\
\llap{$^h$}American Physical Society, One Research Road, Ridge, NY  11961, USA \\
\llap{$^i$}Physics Department, University of the Pacific, Stockton, CA  95211, USA \\
\llap{$^j$}School of Computing, DePaul University, Chicago, IL  60604, USA \\
\llap{$^k$}Department of Physics, University of California, Santa Barbara, CA  93106, USA \\
\llap{$^l$}Department of Physics, Brookhaven National Laboratory, Upton, NY  11973, USA}
\author{The Fermilab Lattice and MILC Collaborations}
\abstract{Lattice calculations of the form factors for the charm semileptonic decays $D\to Kl\nu$ and $D\to\pi l\nu$ provide inputs to direct determinations of the CKM matrix elements $|V_{cs}|$ and $|V_{cd}|$ and can be designed to validate calculations of the form factors for the bottom semileptonic decays $B\to\pi l\nu$ and $B\to Kl\bar{l}$.  We are using Fermilab charm (bottom) quarks and asqtad staggered light quarks on the 2+1 flavor asqtad MILC ensembles to calculate the charm (bottom) form factors.  We outline improvements to the previous calculation of the charm form factors and detail our progress.  We expect our current round of data production to allow us to reduce the theoretical uncertainties in $|V_{cs}|$ and $|V_{cd}|$ from $10.5\%$ and $11\%$, respectively, to about $7\%$.}

\FullConference{The XXVII International Symposium on Lattice Field Theory\\
                 July 26-31, 2009\\
                 Peking University, Beijing, China}

\begin{document}

\section{Introduction}
The CKM matrix elements $|V_{cs}|$ and $|V_{cd}|$ can be extracted to greatest precision (currently to 0.02\% and 0.4\%, respectively) by assuming CKM unitarity and performing a fit to all data~\cite{Amsler:2008zzb}.  However, the simplest tests of unitarity require direct determinations of the CKM matrix elements.  

The decay rate for $D\to K(\pi)l\nu$ is proportional to a form factor and $|V_{cs}|$ ($|V_{cd}|$).  Experiments can measure the decay rates and the form factor shapes, but nonperturbative calculations of the strong force are required to fix the form factor normalizations and extract $|V_{cs(d)}|$.  Therefore these decays allow direct determinations of $|V_{cs(d)}|$ and consistency checks between lattice QCD and unitarity.  Such consistency increases our confidence in both.  

In June CLEO-c published the results of an analysis of $818\ \mr{pb}^{-1}$ collected at charm threshold \cite{Besson:2009uv}.  Combining the CLEO-c results with the first 2+1 flavor lattice calculations of the $D\to K(\pi)l\nu$ form factors~\cite{Aubin:2004ej,Bernard:2009ke} yields $|V_{cs(d)}|$~\cite{Besson:2009uv}:
\begin{eqnarray}
|V_{cs}|&=&0.985(1\pm0.9\%\pm0.6\%\pm10.5\%),\label{cCKM1}\\
|V_{cd}|&=&0.234(1\pm3\%\pm0.9\%\pm11\%).\label{cCKM2}
\end{eqnarray}
The first errors are experimental statistical errors, and the second are experimental systematics.  The third errors are due to uncertainties in the lattice QCD calculations.  The theory errors dominate the uncertainties.  

Discretization effects are the dominant source of the theory errors~\cite{Aubin:2004ej}.  Other uncertainties enter because of incomplete suppression of oscillations due to opposite-parity states, truncation effects in fits to staggered chiral perturbation theory (S$\chi$PT), and model-dependence implicit in the Becirevic-Kaidalov (BK) parameterization~\cite{Aubin:2004ej,Becirevic:1999kt}.  

These sources of uncertainty were addressed in work on $B\to\pi l\nu$ decays~\cite{Bailey:2008wp}.  By calculating the $D\to K(\pi)l\nu$ form factors using the same methods, we may be able to validate their application to calculations of the form factors for $B\to\pi l\nu$ and $B\to Kl\bar{l}$.  The former decay allows a precise determination of $|V_{ub}|$ and a stringent test of unitarity.  The latter is a rare decay and a prime candidate for new physics.  Below we describe our progress in reducing the uncertainties in the charm form factors and anticipate the reduction of the uncertainties in $|V_{cs(d)}|$.

\section{Ensembles and quark masses}
To decrease discretization effects and improve our control of the chiral extrapolation, we are generating full QCD and partially quenched data on each of the ensembles shown in Table~\ref{table:ens}.  These ensembles include the four most chiral coarse ensembles used in the calculations of Ref.~\cite{Aubin:2004ej}, the two fine ($a\approx0.09\ \mr{fm}$) ensembles included in our recent calculation of the form factor for $B\to\pi l\nu$~\cite{Bailey:2008wp}, two additional fine ensembles, three superfine ($a\approx0.06\ \mr{fm}$) ensembles, and one ultrafine ($a\approx0.045\ \mr{fm}$) ensemble~\cite{Bazavov:2009bb}.  The MILC Collaboration has increased the number of configurations in each of the previously used coarse and fine ensembles by a factor of four, and we expect a corresponding decrease in all statistics-dominated uncertainties by a factor of two.  

We have found that randomizing the spatial location of the sources significantly decreases autocorrelations in 2-point functions, which suggests that we may be able to increase our statistics further by increasing the number of source times on each configuration.  We have nearly completed data generation at four source times on the coarse ensembles, the fine ensembles with $m_l=0.4m_s$, $0.2m_s$, and $0.1m_s$, and the superfine ensemble with $m_l=0.2m_s$.  

Power counting arguments~\cite{Aubin:2004ej,Bailey:2008wp} indicate that including these ensembles will effectively eliminate discretization effects due to light quarks and gluons, while heavy-quark discretization effects will be reduced but remain significant.  To improve our estimates of heavy-quark discretization effects, we are investigating including them in chiral-continuum expansions~\cite{Jim}.  This approach incorporates the information from power counting while more systematically fixing the appropriate hadronic scales.  
\begin{table}[tbp]
\centering
\begin{tabular}{lcccclc}
\hline\hline
       & $\approx a$ (fm) & $am_l/am_s$ & Volume & $N_{conf}$ & $am_\mr{valence}$ \\
\hline
coarse & $0.12$ & $0.02/0.05$ & $20^3\times64$ & $2052$ & $0.005,\ 0.007,\ 0.01,$ \\ 
	&       & $0.01/0.05$ & $20^3\times64$ & $2259$ & $0.02,\ 0.03,\ 0.0415,$ \\ 
       	&  	& $0.007/0.05$ & $20^3\times64$ & $2110$ & $0.05;\ 0.0349$	  \\
   	&       & $0.005/0.05$ & $24^3\times64$ & $2099$ &			  \\
\hline
fine   & $0.09$ & $0.0124/0.031$ & $28^3\times96$ & $1996$ & $0.0031,\ 0.0047,\ 0.0062,$\\
	&       & $0.0062/0.031$ & $28^3\times96$ & $1946$ & $0.0093,\ 0.0124,\ 0.031;$	 \\
	&       & $0.00465/0.031$ & $32^3\times96$ & $983$ & $0.0261$	 		\\
     	&   	& $0.0031/0.031$ & $40^3\times96$ & $1015$ &				 \\
\hline
superfine & $0.06$ & $0.0036/0.018$ & $48^3\times144$ & $668$ & $0.0036,\ 0.0072,\ 0.0018,$ \\
          &	   & $0.0025/0.018$ & $56^3\times144$ & $800$ & $0.0025,\ 0.0054,\ 0.0160;$ \\
          &	   & $0.0018/0.018$ & $64^3\times144$ & $826$ & $0.0188$ 		\\
\hline
ultrafine & $0.045$ & $0.0028/0.014$ & $64^3\times192$ & $861$ & TBD \\
\hline\hline
\end{tabular}
\caption{\label{table:ens}Asqtad staggered quark ensembles generated by the MILC Collaboration~\cite{Bazavov:2009bb,Bernard:2001av,Aubin:2004wf} and slated for upcoming heavy-light analyses, together with the valence quark masses being used at each lattice spacing.  The last valence mass listed at each lattice spacing (after the semicolon) is the tuned strange quark mass.  We are presently generating correlators at four source times on each ensemble and investigating the possibility of adding more source times to further increase the total number of source-configurations.}
\end{table}

\section{Correlators and correlator ratios}
The form factors parameterize the hadronic matrix elements of the flavor-changing vector currents,
\begin{equation}
\bra{K(\pi)}V_\mu\ket{D}=\sqrt{2m_D}\left[v_\mu f_{\|}^{D\to K(\pi)}(q^2)+p_{\bot\mu}f_\bot^{D\to K(\pi)}(q^2)\right],\label{ffdef}
\end{equation}
where $V_\mu$ is the lattice current corresponding to $i\bar{s}\gamma_\mu c\ (i\bar{d}\gamma_\mu c)$, $v=p_D/m_D$ is the four-velocity of the $D$ meson, $p_\bot=p_{K(\pi)}-(p_{K(\pi)}\cdot v)v$ is the component of kaon (pion) momentum perpendicular to $v$, and $q^2\equiv(p_D-p_{K(\pi)})^2$ is the invariant mass of the leptons.  We work in the $D$-meson rest frame, in which the form factors are proportional to the temporal and spatial components of the hadronic matrix elements, and $q^2=m_D^2+m_{K(\pi)}^2-2m_DE_{K(\pi)}$.  

One way to extract the hadronic matrix elements is by considering simple ratios of 3-point to 2-point correlators~\cite{Aubin:2004ej},
\begin{equation}
\frac{C_{3,\mu}^{D\to K(\pi)}(t,T;\mb{p}_{K(\pi)})}{C_2^{K(\pi)}(t;\mb{p}_{K(\pi)})C_2^D(T-t)},\label{ratio}
\end{equation}
where $T$ is the separation between source and sink in the 3-point functions, and
\begin{eqnarray}
C_{3,\mu}^{D\to K(\pi)}(t,T;\mb{p}_{K(\pi)})&=&\sum_{\mb{x,y}}e^{i\mb{p}_{K(\pi)}\cdot\mb{y}}\langle\mc{O}_{K(\pi)}(t_i,\mb{0})V_\mu(t,\mb{y})\mc{O}_D^\dag(t_{f},\mb{x})\rangle,\nonumber \\
&&\phantom{\sum_{\mb{x,y}}e^{i\mb{p}_{K(\pi)}\cdot\mb{y}}\langle\mc{O}_{K(\pi)}}t\in[t_i,\ t_{f}=(t_i+T)\;\mr{mod}\;n_t],\\
C_2^{K(\pi)}(t;\mb{p}_{K(\pi)})&=&\sum_\mb{x}e^{i\mb{p}_{K(\pi)}\cdot\mb{x}}\langle\mc{O}_{K(\pi)}(t_i,\mb{0})\mc{O}_{K(\pi)}^\dag(t,\mb{x})\rangle,\nonumber\\
&&\phantom{\sum_\mb{x}e^{i\mb{p}_{K(\pi)}\cdot\mb{x}}\langle\mc{O}_{K(\pi)}}t\in[t_i,\ t_f=(t_i+n_t)\;\mr{mod}\;n_t),\\
C_2^D(t)&=&\sum_\mb{x}\langle\mc{O}_D(t_i,\mb{0})\mc{O}_D^\dag(t,\mb{x})\rangle,\quad t\in[t_i,\ t_f=(t_i+n_t)\;\mr{mod}\;n_t).
\end{eqnarray}
where $n_t$ is the temporal extent of the lattice, and $\mb{p}_{K(\pi)}$ is the momentum of the outgoing kaon (pion).  We calculate the correlators for momenta $\mb{p}_{K(\pi)}=(0,0,0)$, $(1,0,0)$, $(1,1,0)$, $(1,1,1)$, and $(2,0,0)$ (in units of $2\pi/L$, where $L$ is the spatial extent of the lattice) and all times $t$ in the ranges shown.  We increase statistics by averaging correlators with source times $t_i=0,\ n_t/4,\ n_t/2,\ 3n_t/4$.  The $D$-meson interpolating operators $\mc{O}$ are smeared with a charmonium wavefunction to suppress coupling to excited states.  

$C_3$ is calculated with insertions of the current operator at all times $t$ between the source and sink.  At sufficiently large source-sink separations $T$ and times $t$ sufficiently far from both source and sink ($0\ll t\ll T$), a plateau emerges in the ratio~(\ref{ratio}).  This plateau is directly proportional to the desired hadronic matrix element.  

In practice we find that oscillations from opposite-parity excited states contaminate the entire plateau region~\cite{Aubin:2004ej,Bailey:2008wp}.  We therefore consider the more carefully constructed correlator ratios introduced in Ref.~\cite{Bailey:2008wp}:
\begin{equation}
{\overline R}_{3,\mu}^{D\to K(\pi)}(t,T;q^2)\equiv{\frac{1}{\phi_{K(\pi)\mu}}\frac{{\overline{C}}_{3,\mu}^{D\to K(\pi)}(t,T;\mb{p}_{K(\pi)})}{\sqrt {\overline{C}_2^{K(\pi)}(t;\mb{p}_{K(\pi)}){\overline{C}}_2^D(T-t)}}}\sqrt{\frac{2E_{K(\pi)}}{e^{-E_{K(\pi)}t}e^{-m_D(T-t)}}},\label{rat}
\end{equation}
where $\phi_{K(\pi)\mu}\equiv(1,\ \mb{p}_{K(\pi)})$ and the correlators ${\overline C}_3,\ {\overline C}_2$ are constructed from the correlators $C_3,\ C_2$ to eliminate oscillations from opposite-parity states:
\begin{eqnarray}
{\overline C}_3(t,T)&\equiv&\frac{1}{8}\Bigl[C_3(t,T)+C_3(t,T+1)e^{m_D}+2C_3(t+1,T)e^{E_{K(\pi)}-m_D}+2C_3(t+1,T+1)e^{E_{K(\pi)}}\nonumber\\
&+&C_3(t+2,T)e^{2(E_{K(\pi)}-m_D)}+C_3(t+2,T+1)e^{2E_{K(\pi)}-m_D}\Bigr],\label{bar3}\\
{\overline C}_2(t)&\equiv&\frac{1}{4}\left[C_2(t)+2C_2(t+1)e^{m_D}+C_2(t+2)e^{2m_D}\right].\label{bar2}
\end{eqnarray}
Experience suggests that the errors in direct fits to the oscillating states can be larger than errors in simpler fits.  The construction of (\ref{rat}) and (\ref{bar3}, \ref{bar2}) allows us to fit the ratios to constants without introducing systematic errors.  In the plateau region ($0\ll t\ll T$), the ratios ${\overline R}_{3,\mu}^{D\to K(\pi)}$ for $\mu=0$ ($\mu=i$) approach the form factors $f_{\|}^{D\to K(\pi)}$ ($f_\bot^{D\to K(\pi)}$).  

\begin{figure}
\begin{center}
\psfrag{Ratio (arb. norm.)}[b][b][1.2]{${\overline R}_{3,\mu=0}^{D\to\pi}(t,T;q^2)\sim f_{\|}^{D\to\pi}(q^2)$}
{
\includegraphics[width=0.90\textwidth,clip=true]{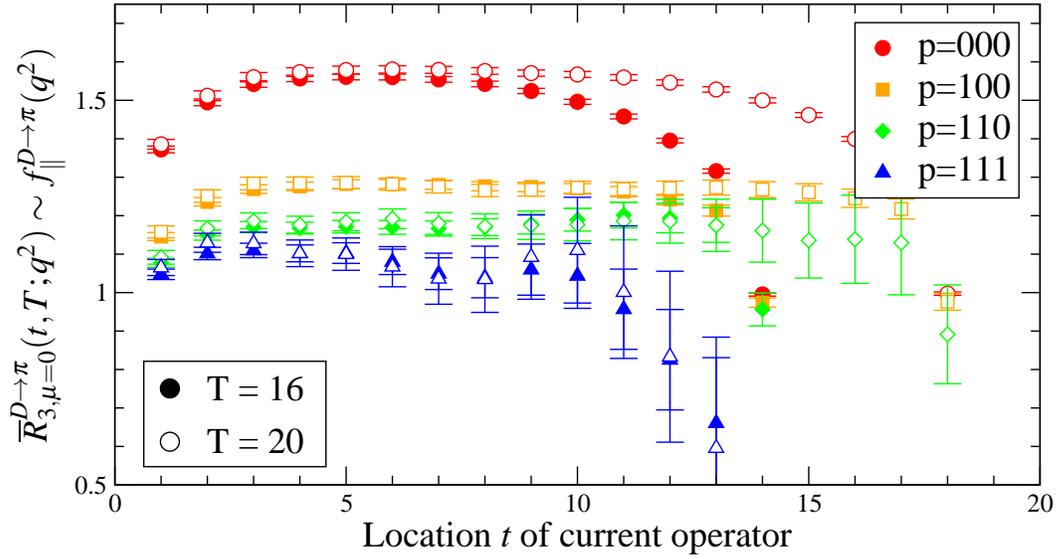}
}
\end{center}
\caption{Ratios of correlators for extracting the form factor $f_{\|}^{D\to\pi}(q^2)$.  The correlators were calculated on $2110$ configurations of the coarse ensemble with $m_l=0.14m_s$.  $T=16,\ 20$ are the source-sink separations, and the three-momenta $\mr{p}$ of the pions are given in units of $2\pi/L$, where $L$ is the spatial extent of the lattice.  Note the excited state contamination in the zero momentum data with $T=16$.}
\label{fig:pi1}
\end{figure}

\begin{figure}
\begin{center}
\psfrag{Ratio (arb. norm.)}[b][b][1.2]{${\overline R}_{3,\mu=i}^{D\to\pi}(t,T;q^2)\sim f_\bot^{D\to\pi}(q^2)$}
{
\includegraphics[width=0.90\textwidth,clip=true]{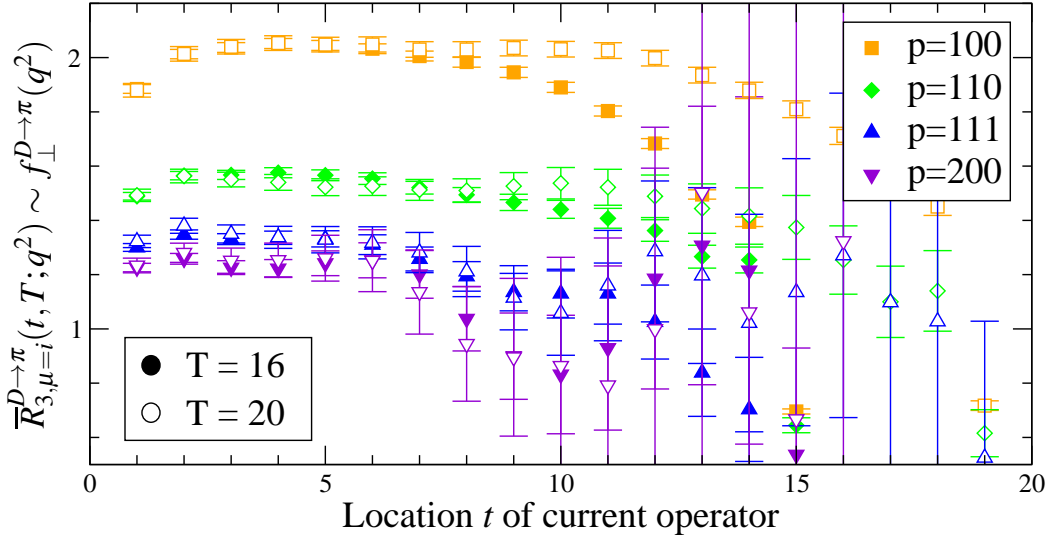}
}
\end{center}
\caption{Ratios of correlators for extracting the form factor $f_\bot^{D\to\pi}(q^2)$.  The correlators were calculated on $2110$ configurations of the coarse ensemble with $m_l=0.14m_s$.  $T=16,\ 20$ are the source-sink separations, and the three-momenta $\mr{p}$ of the pions are given in units of $2\pi/L$, where $L$ is the spatial extent of the lattice.  The consistency of the results for $T=16$ and $T=20$ indicates that the smaller source-sink separation can be used to minimize statistical errors without introducing significant excited state contamination.}
\label{fig:pi2}
\end{figure}

For source-sink separations $T=16$ and $T=20$, examples of the plateaus are shown in Figs.~\ref{fig:pi1} and~\ref{fig:pi2}, where the features leading to the choice of these $T$-values can also be seen.  As the source-sink separation increases, signal-to-noise decreases.  As the source-sink separation decreases, the plateau region shrinks and eventually disappears.  The optimal $T$-value is the smallest for which a plateau exists.  For this $T$, signal-to-noise is maximized without sacrificing the plateau to excited state contamination.  The statistical errors increase with momentum, so the optimal $T$ is momentum dependent.  

To optimize $T$ we generated data with $T=16$, $18$, and $20$ on the coarse $m_l=0.14m_s$ ensemble.  As shown in Figs.~\ref{fig:pi1} and~\ref{fig:pi2}, for $T=20$ plateaus exist for all momenta.  At zero momentum, comparing the $T=16$ data with the $T=20$ data reveals the effects of excited state contamination in the $T=16$ data for all $t$; the plateau has essentially vanished.  At nonzero momentum, comparing the $T=16$ data with the $T=20$ data reveals smaller statistical errors in the $T=16$ data with intact plateau regions.  The larger $T$ allows checks for excited state contamination at smaller momenta, and the smaller $T$ allows us to minimize statistical errors at larger momenta.  On the remaining ensembles, we expect the optimal $T$-value in physical units to be similar.  We are therefore generating data on each ensemble with two $aT$-values of approximately $0.12\ \mr{fm}\times16$ and $0.12\ \mr{fm}\times20$.  
\section{Renormalization and chiral-continuum-energy extrapolation-interpolation}
Lattice form factors obtained from the plateaus in Figs.~\ref{fig:pi1} and \ref{fig:pi2} must be renormalized and extrapolated to zero lattice spacing and the physical light quark masses.  The renormalization factors can be written as products of non-perturbatively calculable factors $Z_V$ and perturbatively calculable factors $\rho$.  The uncertainties in these renormalization factors contribute to the uncertainties in the form factors and CKM matrix elements.  

To perform simultaneous chiral-continuum extrapolations and the kaon (pion) energy interpolation, we can use staggered heavy meson partially quenched chiral perturbation theory ($\chi$PT) with constrained curve fitting~\cite{Bailey:2008wp,Aubin:2005aq,Aubin:2007mc}.  This approach incorporates the energy-dependence of the form factors and yields a model-independent result while accounting for the systematic error due to truncating the expansion.  

To extract $|V_{cs(d)}|$, one can divide the experimental results~\cite{Besson:2009uv} by the lattice form factors evaluated at $q^2=0$.  However, minimizing the uncertainty in $|V_{cs(d)}|$ requires a simultaneous fit to all (experimental and lattice) data.  The analyticity-based parameterization described in Ref.~\cite{Becher:2005bg} captures the energy-dependence of the form factors throughout the kinematic domains, so using it to fit the data and extract CKM matrix elements does not introduce model-dependent systematic errors.  

For $D\to K(\pi)l\nu$, the energy-domains of the lattice and experimental data overlap significantly, allowing a stringent test of the consistency of the shapes of the form factors as determined independently by the lattice and experiment.  This test provides important validation for applying the analyticity-based parameterization to the extraction of $|V_{ub}|$ from $B\to\pi l\nu$, in which the overlap of the lattice data and experimental data is smaller and this self-consistency check, less powerful.  

\section{Expected uncertainties}
A projected error budget for the form factors at $q^2=0$ is shown in Table~\ref{table:D2pi}.  The expected uncertainties reflect previous experience with $B\to\pi l\nu$~\cite{Bailey:2008wp}, including the use of improved correlator ratios, $\chi$PT with constrained curve fitting, and the analyticity-based parameterization to eliminate systematic errors due to incomplete cancellation of oscillating state contributions, truncation of the chiral expansion, and model-dependence in the BK parameterization.  The projections also reflect the four-fold increase in statistics on the coarse ensembles and the addition of the two largest fine ensembles and the superfine $m_l=0.2m_s$ ensemble.  The increase in statistics decreases our statistical uncertainties by a factor of two, while the addition of the superfine ensemble reduces systematic errors due to heavy-quark discretization effects.  
\begin{table}[tbp]
\centering
\begin{tabular}{c|cccccccccc|c|c}
\hline\hline
$\mr{Stat.}+\chi\mr{PT}$ & $g_{D^*D\pi}$ & $r_1$ & $\hat m$ & $m_s$ & $\kappa_c$ & $p_\pi$ & $\mr{HQ}$ & $Z_V$ & $\rho$ & $L^3<\infty$ & $\mr{Sys.}$ & $\mr{Total}$ \\
\hline 
$4.9$ & $2.9$ & $1.4$ & $0.3$ & $1.3$ & $0.2$ & $0.1$ & $3.9$ & $0.7$ & $0.7$ & $0.5$ & $5.4$ & $7.3$ \\
\hline\hline
\end{tabular}
\caption{\label{table:D2pi}Contributions to the relative uncertainties in the form factors at $q^2=0$ assuming data with four source times on the four extended coarse ensembles, two largest fine ensembles, and the superfine $m_l=0.2m_s$ ensemble.  The errors are due to limited statistics and the truncation of chiral perturbation theory; uncertainties in the $D^*D\pi$ coupling, scale, average up-down quark mass, strange quark mass, and charm hopping parameter; momentum-dependent discretization effects from the light quarks and gluons; heavy-quark discretization effects; uncertainties in the renormalization factors $Z_V$ and $\rho$; and finite volume effects.  The last two entries are the total systematics and the total error, both added in quadrature.}
\end{table}

Heavy-quark discretization effects and the uncertainty in the $D^*D\pi$ coupling dominate the systematic uncertainties, while statistics and $\chi$PT truncation error are alone comparable to the entire remaining systematic error.  Heavy-quark discretization effects are sensitive to the smallest lattice spacings included, so they will decrease further with the addition of the ultrafine ensemble in Table~\ref{table:ens}.  The error due to the $D^*D\pi$ coupling may respond to the increased statistics.  From Table~\ref{table:D2pi} and Eqs.~(\ref{cCKM1}) and (\ref{cCKM2}), we expect to reduce the theoretical uncertainties in the CKM matrix elements from about $11\%$ to about $7\%$.  

Fermilab is operated by Fermi Research Alliance, LLC, under Contract No. DE-AC02-07CH11359 with the United States Department of Energy.

\end{document}